\documentclass[epjST]{svjour}
\usepackage{color}
\usepackage{graphics}
\usepackage{epsfig}
\usepackage[latin1]{inputenc}
\usepackage{amssymb}
\usepackage{amsmath}

\graphicspath{{./}{./Figures/}}

\usepackage{latexsym}
\usepackage{mathrsfs}
\usepackage{bm}
\usepackage{comment}

\newcommand{\m}[1]{\begin{pmatrix}#1\end{pmatrix}}

\newcommand{\matr}[1]{{{\bm{#1}}}}    
\renewcommand{\vec}[1]{{\bm{#1}}}    

\begin{document}
\title{Chimera states in complex networks:\\ interplay of fractal topology and delay}
\author{Jakub Sawicki\inst{1}\fnmsep\thanks{\email{zergon@gmx.net}} \and Iryna Omelchenko\inst{1} \and Anna Zakharova\inst{1} \and Eckehard Sch\"{o}ll\inst{1}}
\institute{Institut f\"{u}r Theoretische Physik, Technische Universit\"{a}t Berlin, Hardenbergstr. 36, 10623 Berlin, Germany}
\abstract{
Chimera states are an example of intriguing partial synchronization patterns emerging in networks of identical oscillators. They consist of spatially coexisting domains of coherent (synchronized) and incoherent (desynchronized) dynamics. We analyze chimera states in networks of Van der Pol oscillators with hierarchical connectivities, and elaborate the role of time delay introduced in the coupling term. In the parameter plane of coupling strength and delay time we find tongue-like regions of existence of chimera states alternating with regions of existence of coherent travelling waves. We demonstrate that by varying the time delay one can deliberately stabilize desired spatio-temporal patterns in the system. 
} 
\maketitle
\section{Introduction}
Systems of coupled oscillators are widely studied in the context of nonlinear dynamics, network science, and statistical physics, with a variety of applications in physics, biology, and technology~\cite{PIK01,BOC06a}. Recent increasing interest in such systems is connected with the phenomenon of chimera states~\cite{PAN15,SCH16b}. First obtained in systems of phase oscillators~\cite{KUR02a,ABR04}, chimeras can also be found in a large variety of different systems including time-discrete maps~\cite{OME11,SEM15a,VAD16}, time-continuous chaotic models~\cite{OME12}, neural systems~\cite{OME13,HIZ13,OME15,TSI16}, Boolean networks~\cite{ROS14}, population dynamics~\cite{HIZ15,BAN16}, quantum oscillators~\cite{BAS15}, and in higher spatial dimensions~\cite{OME12a,PAN15,SHI04,MAI15}. Together with the initially reported chimera states, which consist of one coherent and one incoherent domain, new types of these peculiar states having multiple~\cite{OME13,VUE14,OME15a,SET08,XIE14} or alternating~\cite{HAU15} incoherent regions, as well as amplitude-mediated~\cite{SET13,SET14}, and pure amplitude chimera and chimera death states~\cite{ZAK14,BAN15} were discovered. A universal classification scheme for chimera states has recently been proposed~\cite{KEM16}.

Chimera states account for numerous applications in natural and technological systems, such as uni-hemispheric sleep~\cite{RAT00,RAT16}, bump states in neural systems~\cite{LAI01,SAK06a}, epileptic seizures~\cite{ROT14,AND16}, power grids~\cite{MOT13a}, or social systems~\cite{GON14}. 
Experimentally, chimeras have been found in optical~\cite{HAG12}, chemical~\cite{TIN12,NKO13} systems, mechanical~\cite{MAR13,KAP14}, electronic~\cite{LAR13,GAM14}, optoelectronic delayed-feedback~\cite{LAR15} and electrochemical~\cite{WIC13,SCH14a} oscillator systems, Boolean networks~\cite{ROS14}, and optical combs~\cite{VIK14}.

Recent studies have shown that not only nonlocal coupling schemes, but also global~\cite{SET14,YEL14,BOE15,SCH15a,SCH15e}, as well as more complex coupling topologies allow for the existence of chimera states~\cite{TSI16,HIZ15,KO08,OME15,ULO16}. Furthermore, time-varying network structures can give rise to alternating chimera states~\cite{BUS15}. Chimera states have also been shown to be robust against inhomogeneities of the local dynamics and coupling topology~\cite{OME15}, against noise~\cite{LOO16}, or they might be even induced by noise~\cite{SEM15b,SEM16,ZAK17}. 

An interesting example of complex network topology are networks with hierarchical connectivities, arising in neuroscience as a result of Diffusion Tensor Magnetic Resonance Imaging analysis, showing that the connectivity of the neuron axons network represents a hierarchical (quasi-fractal) geometry~\cite{KAT09,EXP11,KAT12,KAT12a,PRO12}.
Such network topology can be realized using a Cantor algorithm starting from a chosen base pattern~\cite{OME15,ULO16}, and is in the focus of our study in the present manuscript. 

Current analysis of chimera states in oscillatory systems has demonstrated possible ways to control chimera states~\cite{SIE14c,BIC15,OME16}, extending their lifetime and fixing their spatial position. It is well known that time delay can also serve as an instrument for stabilization/destabilization of complex patterns in networks. 

It is worth mentioning here that networks of coupled oscillators with complex topologies are often characterized by high multistability, which makes the investigation of different complex spatio-temporal patterns a challenging problem. The goal of the present study is to study the influence of time delay on chimera states in networks of Van~der~Pol oscillators with hierarchical connectivity, and to demonstrate how by varying the time delay one can stabilize chimera states in the network.

\section{The Model}
\label{sec:model}
We consider a ring of $N$ identical Van der Pol oscillators with different coupling topologies, which are given by the respective adjacency matrix $\matr{G}$. While keeping the periodicity of the ring, and the circulant structure of the adjacency matrix, we vary the connectivity pattern of each element. The dynamical equations for the 2-dimensional phase space variable $\vec{x}_k=(u_k, \dot{u}_k)^T=(u_k, v_k)^T \in \mathbb{R}^2$ are:
\begin{align}
\vec{\dot{x}}_i(t) &=  \vec{F}(\vec{x}_i(t)) +  \frac{\sigma}{g} \sum^{N}_{j=1}G_{ij} \matr{H}[\vec{x}_j(t-\tau)-\vec{x}_i(t)]
\label{eqn:gen1}
\end{align}
with $i \in \{1,...,N\}$ and the delay time $\tau$. The dynamics of each individual oscillator is governed by 
\begin{eqnarray}
\label{eq:localdyn}
\vec{F}(\vec{x})=
\left(\!
\begin{array}{*{1}{c}}
v\\
\varepsilon(1-u^2)v-u
\end{array}
\!\right) ,
\end{eqnarray}
where  $\varepsilon$ denotes the bifurcation parameter.
The uncoupled Van der Pol oscillator has a stable fixed point at $\vec{x}=0$ for $\varepsilon<0$ and undergoes an Andronov-Hopf bifurcation at $\varepsilon=0$. Here, only $\varepsilon=0.1$ is considered.  The parameter~$\sigma$ denotes the coupling strength,  and $g=\sum^{N}_{j=1}G_{ij}$ is the number of links for each node (corresponding to the row sum of $\matr{G}$). The interaction is realized through diffusive coupling with coupling matrix $\matr{H}=\m{0&0\\b_1&b_2}$ and real interaction parameters $b_1$ and $b_2$. In accordance with Omelchenko et al.~\cite{OME15a}, throughout the manuscript we fix the parameters $b_1=1.0$ and $b_2=0.1$.

\subsection{Fractal topology}
Fractal topologies can be generated using a classical Cantor construction algorithm for a fractal set~\cite{MAN83,FED88}. This iterative hierarchical procedure starts from a \emph{base pattern} or initiation string $b_{init}$ of length $b$, where each element represents either a link ('$1$') or a gap ('$0$'). The number of links contained in $b_{init}$ is referred to as $c_1$. In each iterative step, each link is replaced by the initial base pattern, while each gap is replaced by $b$ gaps. Thus, each iteration increases the size of the final bit pattern, such that after $n$ iterations the total length is $N=b^n$. We call the resulting pattern fractal. 
Using the resulting string as the first row of the adjacency matrix $\matr{G}$, and constructing a circulant adjacency matrix $\matr{G}$ by applying this string to each element of the ring, a ring network of $N=b^n$ nodes with hierarchical connectivity is generated ~\cite{OME15,HIZ15,TSI16}. Here we slightly modify this procedure by including an additional zero in the first instance of the sequence, which corresponds to the delayed self-coupling. Therefore, there is no net effect of the diagonal elements of the adjacency matrix $G_{ii}$ on the network dynamics, and hence the first link in the clockwise sense from the reference node is effectively removed from the link pattern. Without our modification, this would lead to a breaking of the base pattern symmetry, i.e., if the base pattern is symmetric, the resulting coupling topology would not be so, since the first link to the right is missing from the final link pattern.
Our procedure, in contrast, ensures the preservation of an initial symmetry of $b_{init}$ in the final link pattern, which is crucial for the observation of chimera states, since asymmetric coupling leads to a drift of the chimera ~\cite{BIC15,OME16}. Thus, a ring network of $N=b^n+1$ nodes is generated.

\subsection{Chimera states in fractal topologies}

Throughout this manuscript, we consider the network generated with base pattern $b_{init}=(11011)$ after four iterative steps. This results in a ring network of $N=5^4 + 1= 626$ nodes. Our choice is motivated by previous studies of chimera states in nonlocally coupled networks~\cite{OME13,OME15a} and networks with hierarchical connectivity~\cite{OME15,ULO16}. In the first case, it has been shown, that an intermediate range of coupled neighbours is crucial for the observation of chimera states, too large and too small numbers of connections makes this impossible. In the second case, it has been demonstrated that hierarchical networks with higher clustering coefficient promote chimera states. Exploiting the clustering coefficient $C$ introduced by Watts and Strogatz~\cite{WAT98}, we obtain for the fractal topology a value of $C=0.428$.

\begin{figure}[htbp]
\centering
\includegraphics[width = 1.0\textwidth]{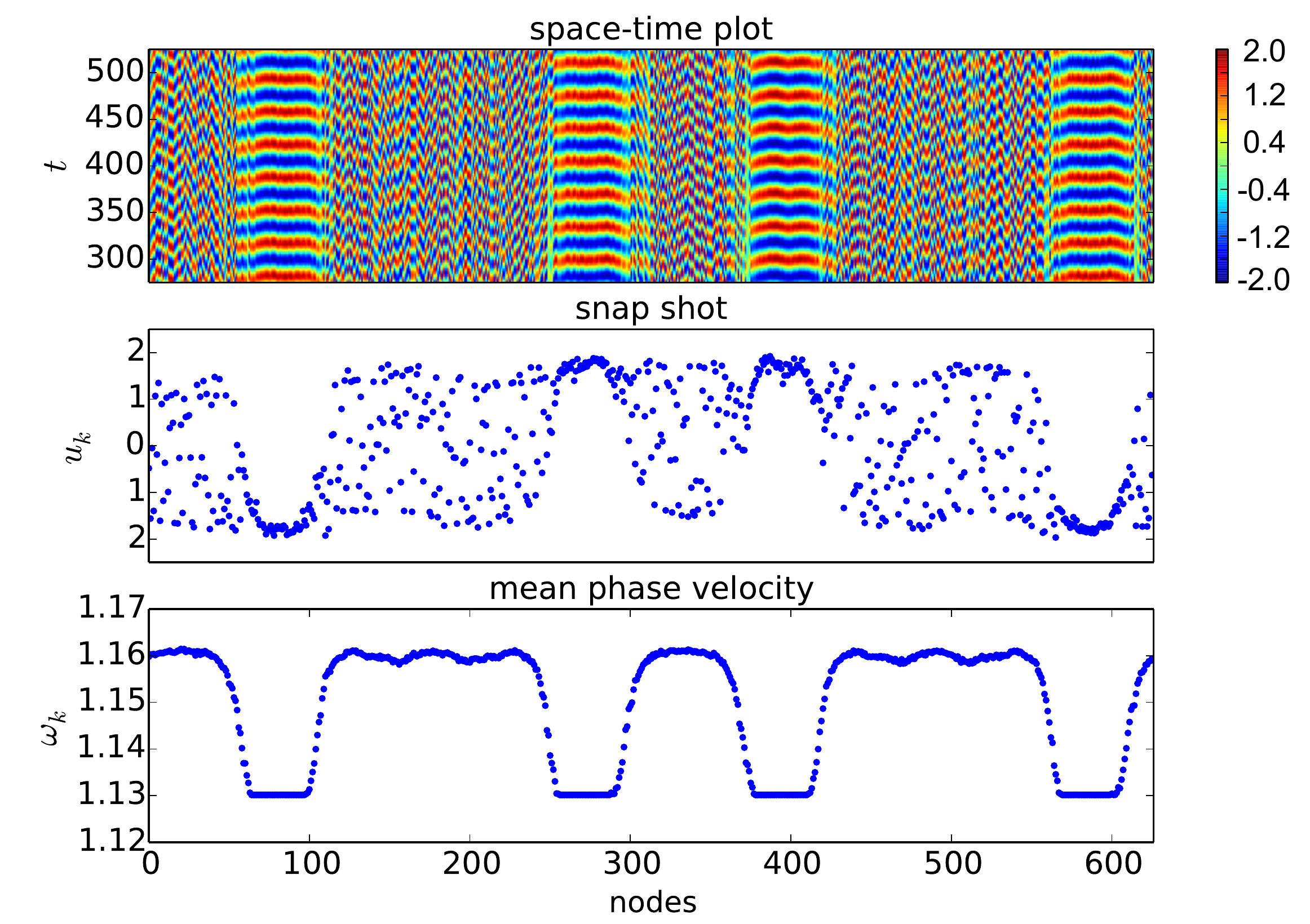}
\caption{(Color online) Chimera state in the undelayed case $\tau = 0$ for  $b_{init}=(11011)$, $n=4$, $N=626$, $\varepsilon=0.1$, and $\sigma=0.35$. Note the nonidentical sizes of incoherent domains. The three panels correspond to the same simulation: 
Space-time plot of $u$ (upper panels), snapshots of variables $u_k$ at $t=1000$ (middle panels), and mean phase velocity profile $\omega_k$ (bottom panels). This asymmetric pattern is used as initial condition for further simulations with $\tau \ne 0$.
}
\label{fig:1}
\end{figure}

\section{Influence of time delay}

Figure\,\ref{fig:1} demonstrates chimera states in the system~(\ref{eqn:gen1}) for $b_{init}=(11011)$, $n=4$, $N=626$, $\varepsilon=0.1$, and $\sigma=0.35$, without time delay~$\tau=0$, obtained numerically for symmetric chimera-like initial conditions. We analyze space-time plot (upper panel), the final snapshot of variables $u_i$ at $t=1000$ (middle panel), and frequencies of oscillators averaged over time window $\Delta T = 10000$ (bottom panel). Oscillators from coherent domains are phase-locked and have equal mean frequencies. Arc-like profiles of mean frequencies for oscillators from incoherent domain are typical for chimera states.     

To uncover the influence of time delay introduced in the coupling term in system~(\ref{eqn:gen1}), we analyze numerically the parameter plane of coupling strength~$\sigma$ and delay time~$\tau$. Fixing network parameters $b_{init}=(11011)$, $n=4$, $N=626$, and $\varepsilon=0.1$, we choose the chimera pattern of the undelayed system (shown in Fig.\,\ref{fig:1}) as an initial condition, and vary the values of $\sigma$ and $\tau$. In numerical simulations of chimera states, the choice of initial conditions often plays a very important role. Usually, chimera states coexist with the fully synchronized state or coherent traveling waves, and random initial conditions rarely result in chimera patterns. In contrast, specially prepared initial conditions which combine coherent and incoherent spatial domains, increase the probability of observing chimeras. Nevertheless, it is remarkable that the asymmetric structure in Fig.\,\ref{fig:1} evolves from symmetric initial conditions.

\begin{figure}[htbp]
\centering
\includegraphics[width = 1.05\textwidth]{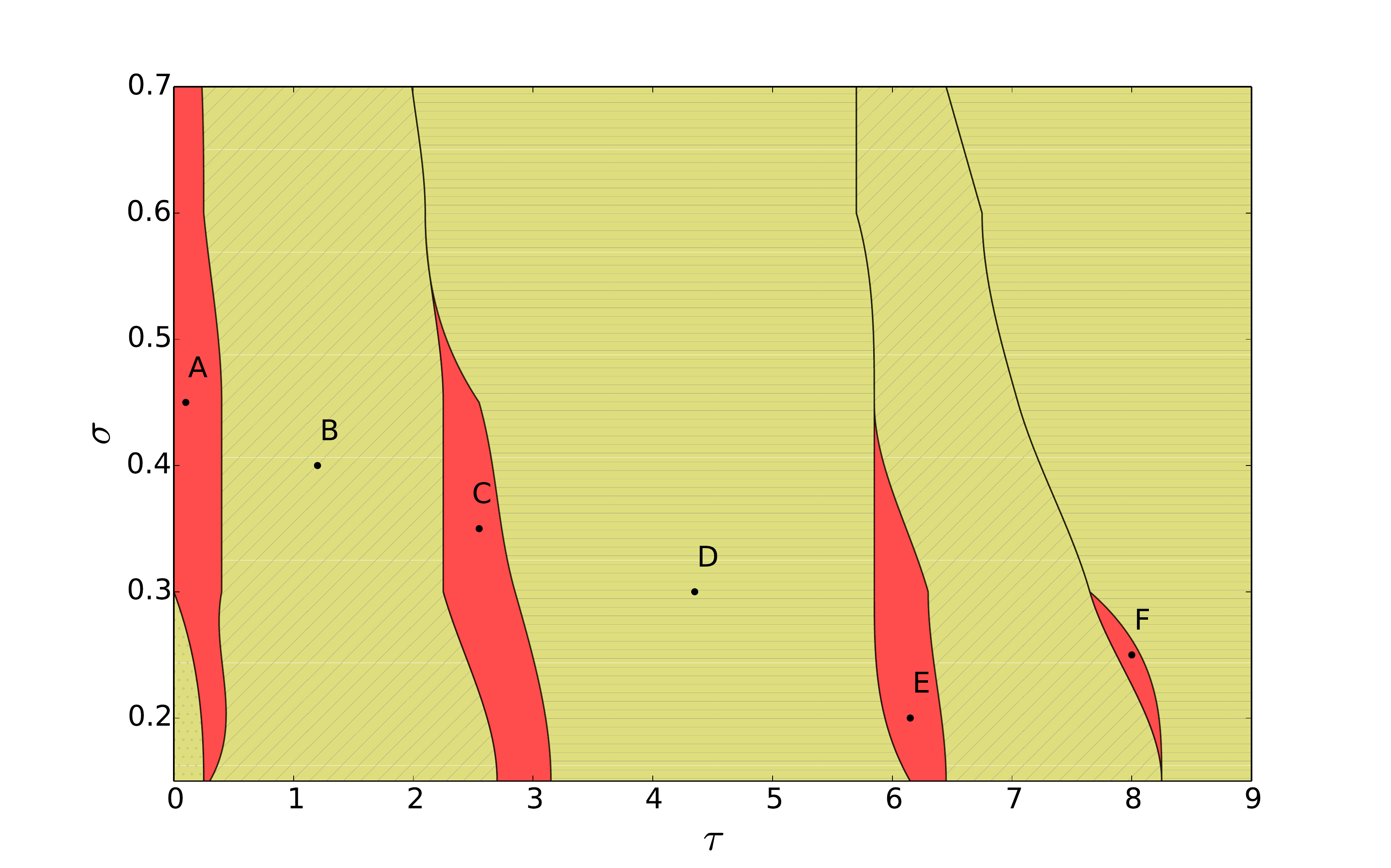}
\caption{(Color online) Chimera tongues (red), in-phase synchronization (horizontally striped yellow region) and coherent traveling waves (diagonally striped yellow region) in the parameter plane~$(\tau,\sigma)$ for $b_{init}=(11011)$, $n=4$, $N=626$, $\varepsilon=0.1$. At the transition to a chimera region we can observe chaos (dotted yellow region at small $\tau,\sigma$).
}
\label{fig:2}
\end{figure}

Fig.\,\ref{fig:2} demonstrates the map of regimes in the parameter plane~$(\tau,\sigma)$. In the undelayed case $\tau=0$ we observe the chimera state shown in Fig.\,\ref{fig:1}. The introduction of small time delay for weak coupling strength immediately destroys the chimera pattern and the incoherent domains characterized by chaotic dynamics appear (yellow dotted region). Nevertheless, for larger values of coupling strength~$\sigma$ chimera states are still present. With increasing delay $\tau$ we observe a sequence of tongue-like regions (shown red) for chimera states. These regions appear in between large areas of alternating coherent structures: fully synchronized states (yellow regions with horizontal stripes) and traveling waves (yellow regions with diagonal stripes). Closer inspection of the chimera tongues shows that increasing $\tau$ reduces the size of the tongues, and also decreases the maximal~$\sigma$ values, for which chimera states are observed. Moreover, one can easily see that chimera regions appear at $\tau$ values close to integer multiples of~$\pi$. 

\begin{figure}[htbp]
\centering
\includegraphics[width = 0.494\textwidth]{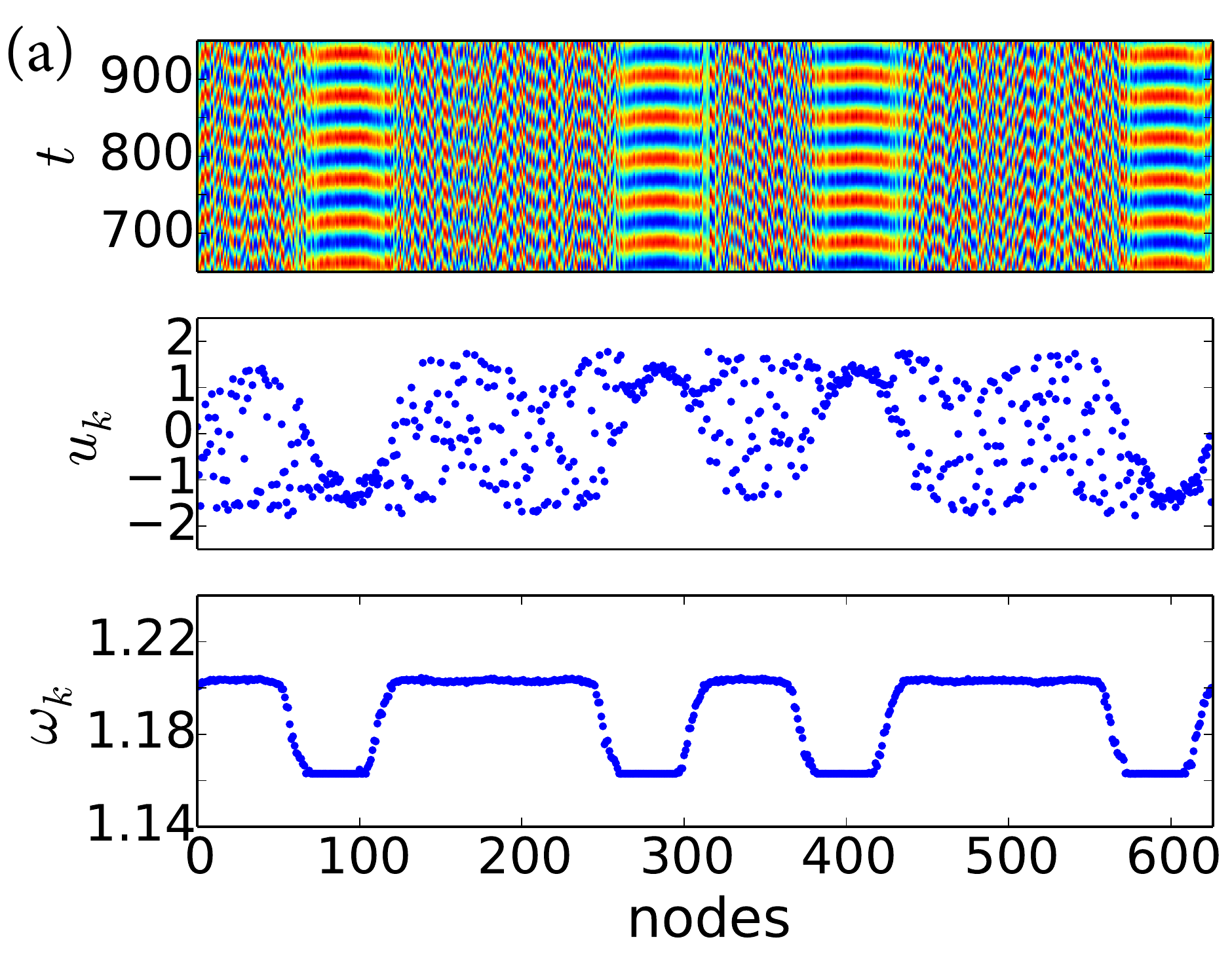}
\includegraphics[width = 0.494\textwidth]{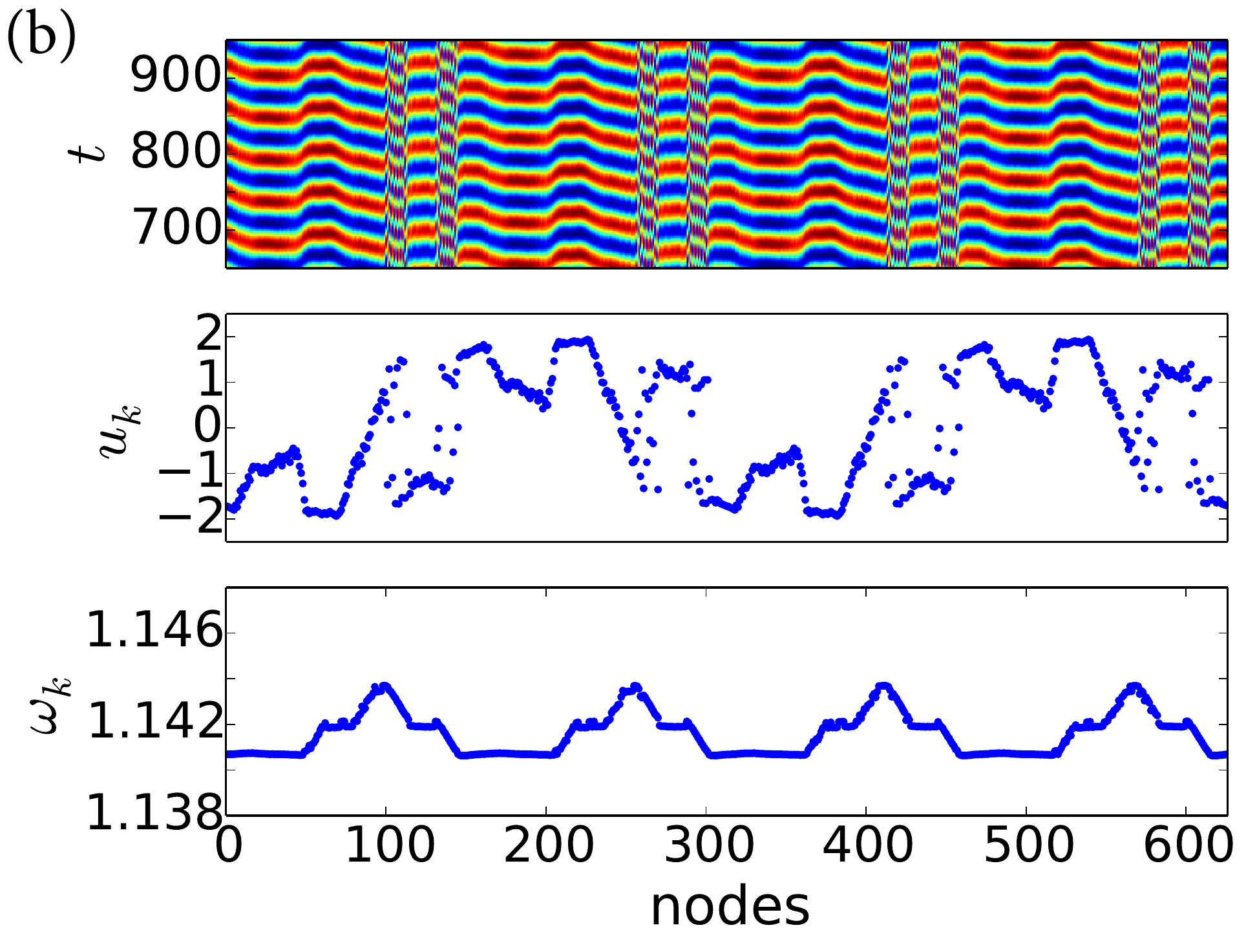}
\includegraphics[width = 0.494\textwidth]{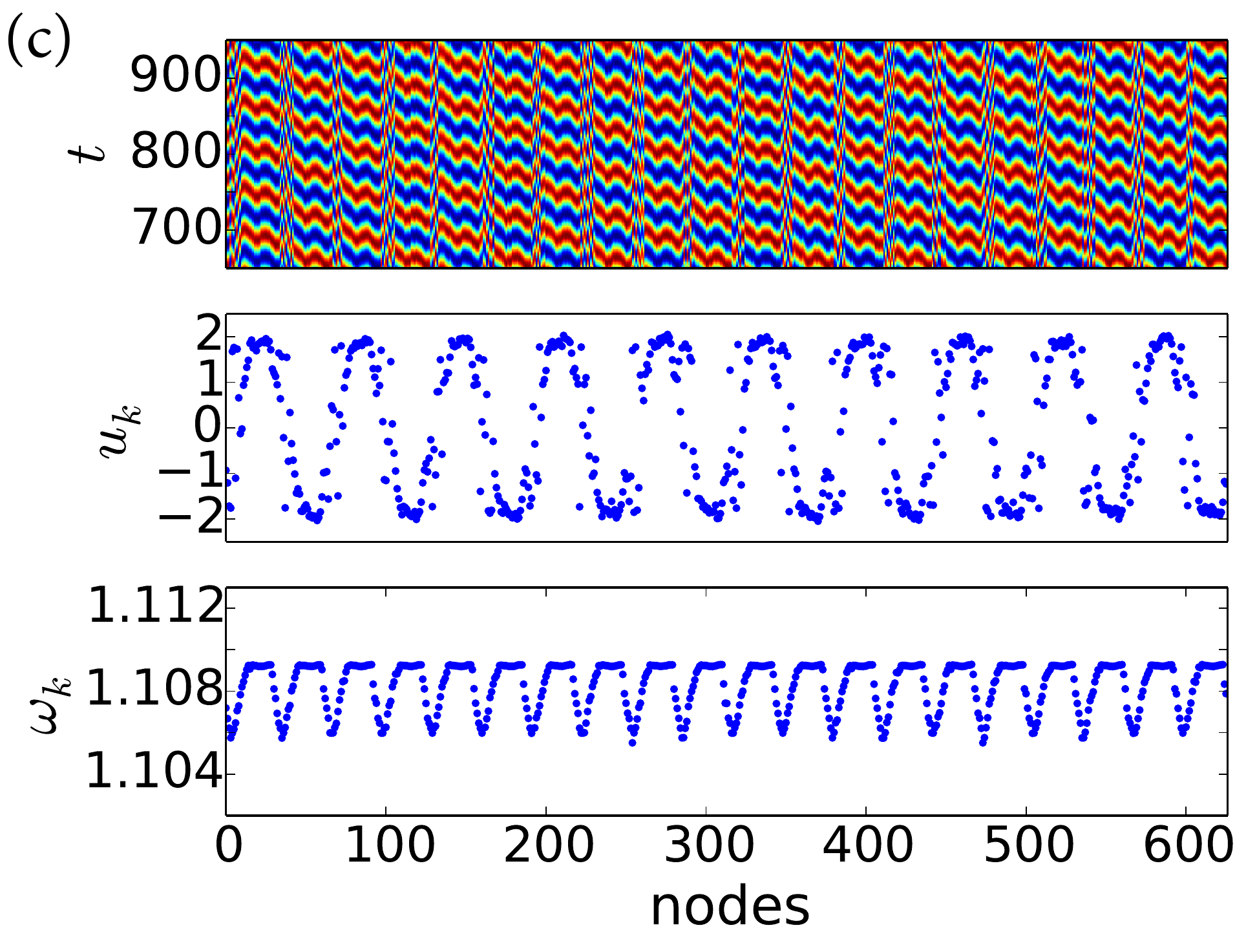}
\includegraphics[width = 0.494\textwidth]{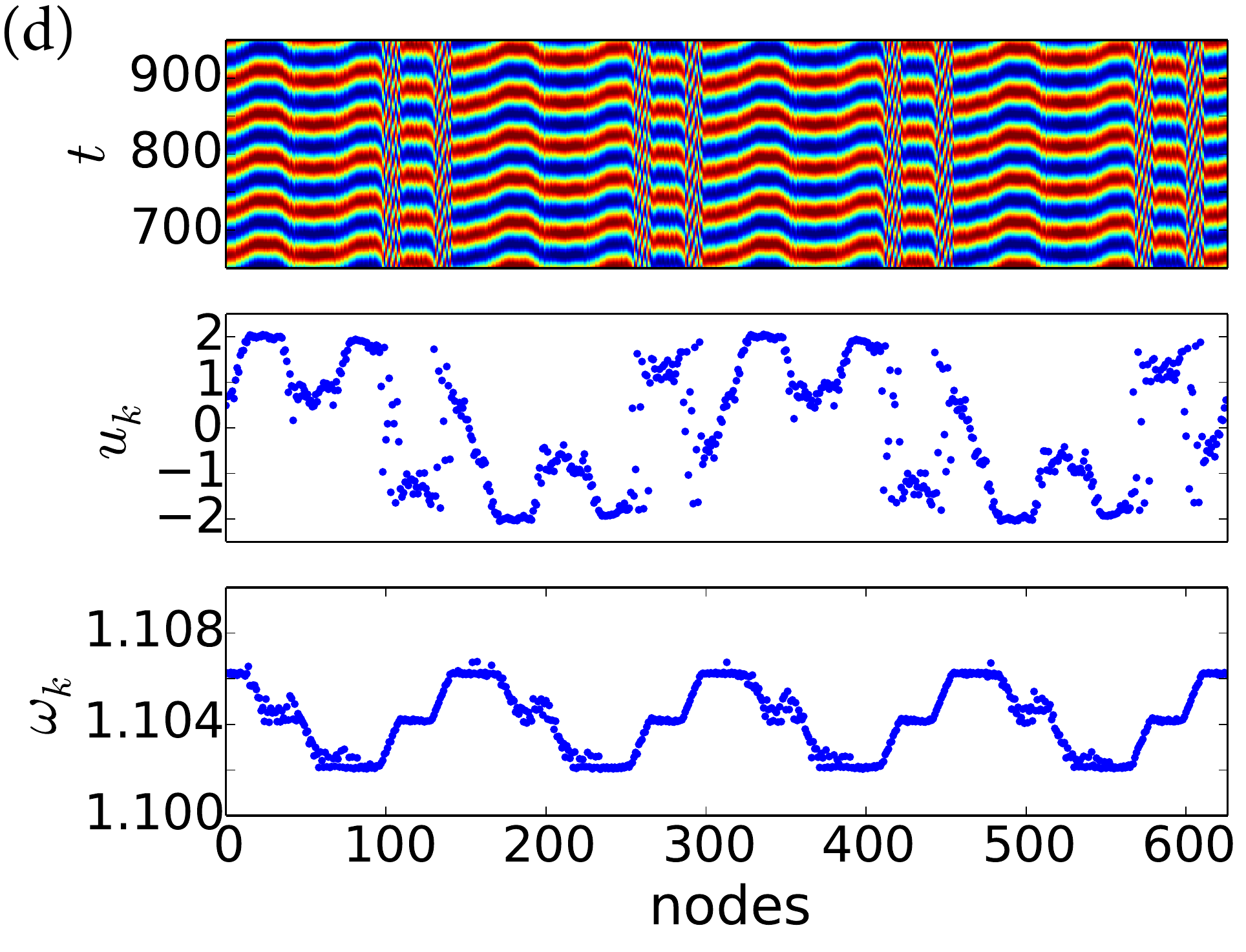}
\caption{(Color online) Patterns taken from the chimera tongues in Fig.\,\ref{fig:2} with $b_{init}=(11011)$, $n=4$, $N=626$, $\varepsilon=0.1$: Space-time plot of $u$ (upper panels), snapshots of variables $u_k$ (middle panels), and mean phase velocity profile $\omega_k$ (bottom panels) for (a) $\tau = 0.1$ and $\sigma=0.45$ (point A), (b) $\tau = 2.55$ and $\sigma=0.35$ (point C), (c) $\tau = 6.15$ and $\sigma=0.20$ (point E), and (d) $\tau = 8.1$ and $\sigma=0.25$ (point F).}
\label{fig:3}
\end{figure}

The sequence of tongues for chimera states in the $(\tau,\sigma)$ parameter plane of system~(\ref{eqn:gen1}) shown in Fig.\,\ref{fig:2} can be understood as a resonance effect in $\tau$~\cite{HOE05,YAN06}. The intrinsic frequency of the uncoupled system is $\omega = 1$ which corresponds to a period of $2 \pi$. Due to the influence of the coupling term the period decreases (see Fig.\,\ref{fig:3} upper panels), therefore, chimera tongues are shifted to the left for increasing coupling strength $\sigma$. 

Let us take a closer look at the dynamics inside the tongues. For the parameter values chosen inside the first, leftmost and largest, tongue we find chimera states similar to the initial condition in Fig.\,\ref{fig:1}. In the second and the forth tongue nested chimera structures can be observed (see Fig.\,\ref{fig:3}b and d). In the third tongue for $\tau \approx 2\pi$ multichimera states can be observed, e.g., a $20-$chimera in Fig.\,\ref{fig:3}c. Therefore, the appropriate choice of time delay $\tau$ in the system allows one to achieve the desired chimera pattern.

\begin{figure}[htbp]
\centering
\includegraphics[width = 0.494\textwidth]{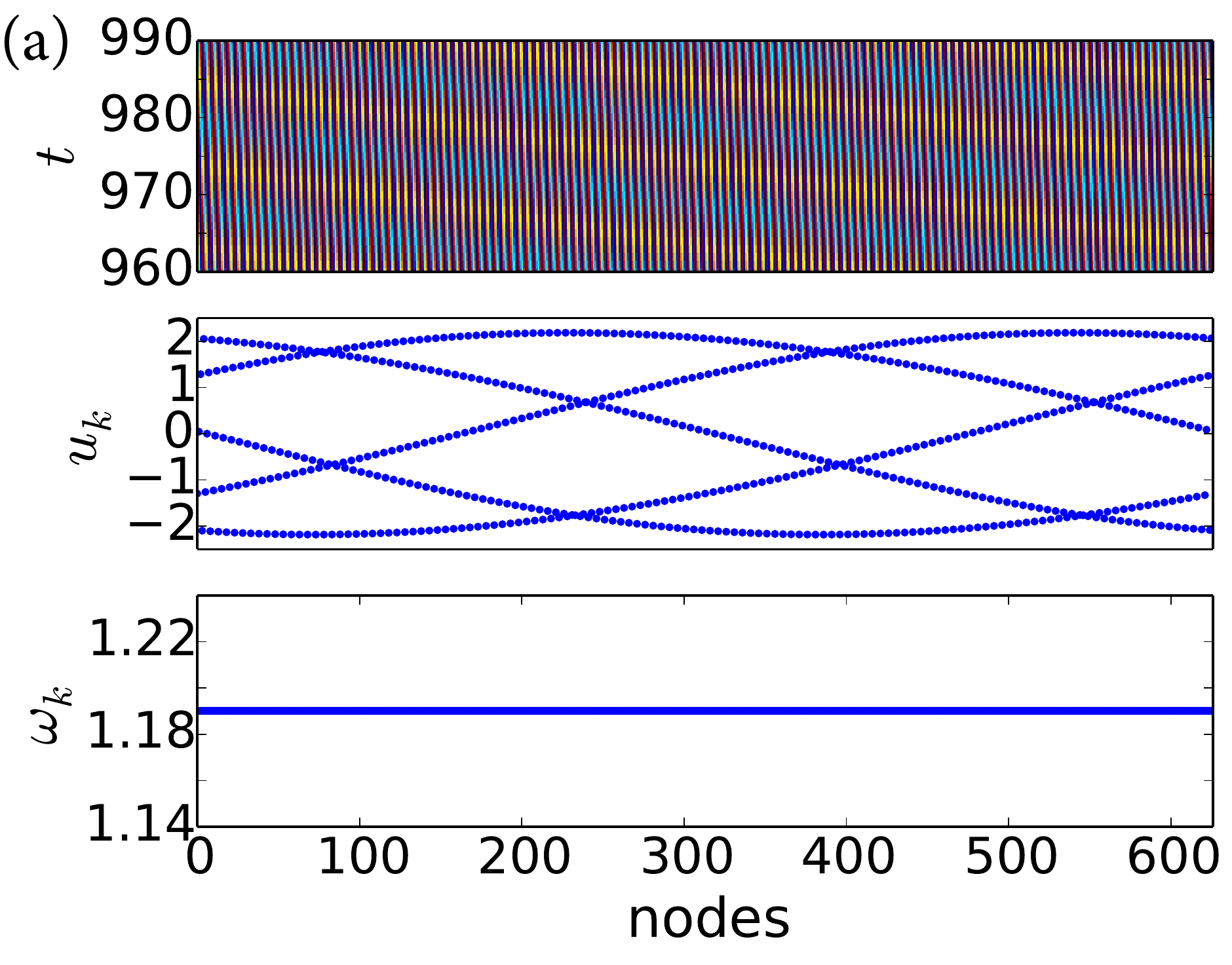}
\includegraphics[width = 0.494\textwidth]{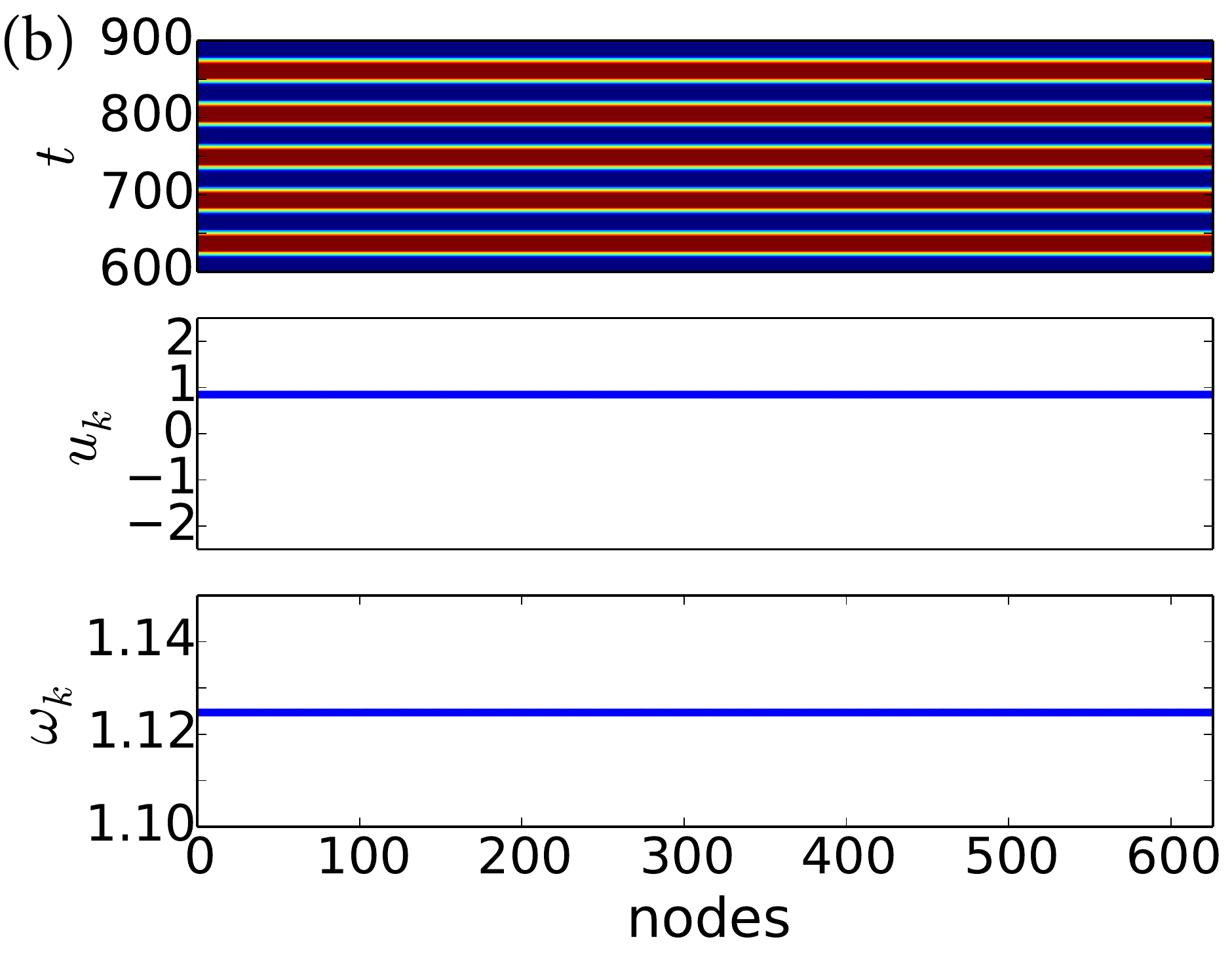}
\caption{(Color online) Patterns taken from the coherent (yellow) regions in Fig.\,\ref{fig:2} with $b_{init}=(11011)$, $n=4$, $N=626$, $\varepsilon=0.1$: Space-time plot of $u$ (upper panels), snapshots of variables $u_k$ (middle panels), and mean phase velocity profile $\omega_k$ (bottom panels) for (a) $\tau = 1.20$ and $\sigma=0.4$ (point B), and (b) $\tau = 4.35$ and $\sigma=0.3$ (point D).
}
\label{fig:4}
\end{figure}

In the parameter plane of delay time and coupling strength the region corresponding to coherent states is dominating (yellow regions in Fig.\,\ref{fig:2}). On one hand, we observe the in-phase synchronization regime (see Fig.\,\ref{fig:4}b) which is enlarged for increasing coupling strength. On the other hand, we also detect a region of coherent traveling waves with wavenumber $k>1$ (see Fig.\,\ref{fig:4}a). Varying the delay time~$\tau$ allows not only for switching between these states, but also for controlling the speed of traveling waves: in the diagonal striped region in Fig.\,\ref{fig:2} the mean phase velocity decreases for increasing delay times. The pyramidal structure of the mean phase velocity profile in Figs.\,\ref{fig:3}b, d is due to the fact that the whole chimera structure is travelling. The speed of travelling is sensitive to the coupling strength and delay time. For a pronounced profile of the mean phase velocity this speed must be small. Otherwise it is smeared out over time.

\section{Discussion}

In the current study, we have analyzed chimera states in ring networks of Van der Pol oscillators with hierarchical connectivities.  
For a fixed base pattern, we have constructed a hierarchical connectivity, and provided a numerical study of complex spatio-temporal patterns in the network. Our study was focused on the role of time delay in the coupling term and its influence on the chimera states.

In the parameter plane of time delay~$\tau$ and coupling strength~$\sigma$, we have determined the stability regimes for different types of chimera states, alternating with regions of coherent states. An appropriate choice of time delay allows us to stabilize several types of chimera states. The interplay of complex hierarchical network topology and time delay results in a plethora of patterns: we observe chimera states with coherent and incoherent domains of non-identical sizes and non-equidistantly distributed in space. Moreover, traveling and non-traveling chimera states can be obtained for a proper choice of time delay. We also demonstrate that time delay can induce patterns which are not observed in the undelayed case.

Our analysis has shown that networks with complex hierarchical topologies, as arising in neuroscience, can demonstrate many nontrivial patterns. Time delay can play the role of a powerful control mechanism which allows either to promote or to destroy chimera patterns.

\section{Acknowledgment}
This work was supported by DFG in the framework of SFB 910. 

\bibliographystyle{epj}

\end{document}